\documentclass{acm_proc_article-sp}

\usepackage{latexsym}
\usepackage{booktabs}
\usepackage{multirow}
\usepackage{float}

\usepackage{epstopdf}

\floatstyle{plaintop}
\restylefloat{table}
\usepackage[tight]{subfigure}
\usepackage{caption}
\usepackage[ruled, lined, linesnumbered, commentsnumbered]{algorithm2e}

 \newtheorem{definition}{Definition}[section]
\newtheorem{theorem}{Theorem}
\newdef{example}{Example}

\newtheorem{lemma}{Lemma}

\newcommand{\squishlist}{
\begin{list}{$\bullet$}
{ \setlength{\itemsep}{0pt} \setlength{\parsep}{3pt}
\setlength{\topsep}{3pt} \setlength{\partopsep}{0pt}
\setlength{\leftmargin}{1.5em} \setlength{\labelwidth}{1em}
\setlength{\labelsep}{0.5em} } }

\newcommand{\squishend}{
\end{list}  }

\newcounter{Lcount}
\newcommand{\squishlisttwo}{
\begin{list}{\arabic{Lcount}.}
{ \usecounter{Lcount}
\setlength{\itemsep}{0pt}
\setlength{\parsep}{0pt}
\setlength{\topsep}{0pt}
\setlength{\partopsep}{0pt}
\setlength{\leftmargin}{2em}
\setlength{\labelwidth}{1.5em}
\setlength{\labelsep}{0.5em} } }

\newcommand{\squishlisttwob}{
\begin{list}{$\bullet$}
{ \setlength{\itemsep}{0pt} \setlength{\parsep}{0pt}
\setlength{\topsep}{0pt} \setlength{\partopsep}{0pt}
\setlength{\leftmargin}{2em} \setlength{\labelwidth}{1.5em}
\setlength{\labelsep}{0.5em} } }

\begin{document}

\title{{\text SharesSkew:} An Algorithm to Handle Skew for Joins in MapReduce}

%
%
%
%
%

\numberofauthors{4} 
\author{
%
%
\alignauthor
Foto Afrati\\
      \affaddr{National Technical University of Athens}\\
      \affaddr{Greece}\\
      \email{afrati@gmail.com}
\alignauthor
Nikos Stasinopoulos\\
      \affaddr{National Technical University of Athens}\\
      \affaddr{Greece}\\
      \email{nstasinopoulos@gmail.com}
\alignauthor
Jeffrey D.~Ullman\\
      \affaddr{Stanford University}\\
      \affaddr{USA}\\
      \email{ullman@gmail.com}
\and  
\alignauthor
Angelos Vassilakopoulos\\
      \affaddr{National Technical University of Athens}\\
      \affaddr{Greece}\\
      \email{avasilako@gmail.com}
}

\maketitle
\begin{abstract}
In this paper, we investigate the problem of computing a multiway join in one round of MapReduce when the data may be skewed.  We optimize on communication cost, i.e., the amount of data that is transferred from the mappers to the reducers.
We identify join attributes values that appear very frequently, Heavy Hitters (HH).
We distribute HH valued records to reducers avoiding skew by using an adaptation of the Shares~\cite{AfUl} algorithm  to achieve  minimum communication cost.
Our algorithm is implemented for experimentation and is offered as open source software.
Furthermore, we investigate a class of multiway joins for which a simpler variant of the
algorithm can handle skew. We offer closed forms for computing the parameters of the algorithm for
chain and symmetric joins.

\end{abstract}
%
%

\section{Introduction}
We study data skew that occurs when we want to compute a multiway join in a single MapReduce round.
When the map phase produces key-value pairs, some keys may receive a significant overload because
many tuples having the same value  on a specific attribute ({\em heavy hitter}) may be present in the data.
On the other hand, it is well recognized that
in MapReduce, the shuffle phase may add significant overhead if we are not careful with how we distribute the inputs even in the case when we see that all keys receive almost the same amount of inputs~\cite{AfratiBCPU11,BeameKS13,jeff}.
The overhead of the shuffle phase depends on the {\em communication cost} which is the amount of data transferred from the mappers to the reducers.
In this paper we develop an algorithm which handles skew in a way that minimizes the communication cost.

The algorithm assumes a preliminary round that identifies the heavy hitters (HH). Then it decomposes the given join in a set of {\em residual } joins, each of which is the original join with two differences:
\begin{itemize}
\item It is applied on a different piece of the data (essentially the tuples that contain the HH values that identify the residual join).
\item The map function is different (to minimize the communication cost under the (different) constraints on sizes of (parts of) relations involved in the residual join).
\end{itemize}
For each residual join, we use the Shares algorithm (\cite{AfUl}) to compute and minimize the communication cost.


The paper is structured as follows:
 In the rest of this section we explain SharesSkew algorithm on 2-way join and, in the end of the section, we explain our formal setting.
 In Section 2, related work can be found.
 In Section 3, an overview of the Shares algorithm from \cite{AfUl} is presented.
 In Section 4, we give an overview of SharesSkew algorithm and we relate the reducer size (which is a parameter that sets a bound on the number of inputs a reducer can receive  and controls the degree of parallelization in the algorithm) and the number of reducers.
 In Section 5, we present the algorithm SharesSkew and in Section 6 we give an extended example of how to apply this algorithm. Section 7 explains how efficiency of SharesSkew is a consequence of the efficiency of Shares algorithm.
 Then, in Section 8, we give closed forms for shares (see Section 2 for their definition) and communication cost for chain joins and symmetric joins.
%
 Finally, Section 9 contains the experiments which verify our analysis about
the performance of SharesSkew algorithm.


\subsection{Algorithm SharesSkew on 2-way Join}
Suppose we have the join $R(A,B) \bowtie S(B,C)$.
Systems such as Pig or Hive that implement SQL or relational algebra over MapReduce have mechanisms to deal with joins where there is significant skew
(see, e.g.,~\cite{hive,pig,Gates09}).
These systems use a two-round algorithm, where the first round identifies the heavy hitters. In the second round, tuples that do not have a heavy-hitter for the join attribute(s) are handled normally. That is, there is one reducer\footnote{In this paper, we use the term {\em reducer} to denote the application of the Reduce function to a key and its associated list of values. It should not be confused with a Reduce task, which typically executes the Reduce function on many keys and their associated values.}
for each key, which is associated with a value of the join attribute.
 Since the key is not a heavy hitter, this reducer handles only a small fraction of the tuples, and thus will not cause a problem of skew.
For tuples with heavy hitters,  new keys are created that are handled along with the other keys (normal or those for other heavy hitters) in a single MapReduce job.
The new keys in
these systems are created with  a simple technique as in the following example:
\begin{example}
\label{1-ex}
We have to compute the join $R(A,B) \bowtie S(B,C)$ using a given number, $k$, of reducers.
Suppose  value $b$ for attribute $B$ is identified as a heavy hitter.
Suppose there are $r$ tuples of $R$ with $B=b$ and there are $s$ tuples of $S$ with $B=b$. Suppose also for convenience that $r>s$.
The distribution to $k$ buckets/reducers is done in earlier approaches by partitioning the data of one of the relations in $k$ buckets (one bucket for each reducer) and sending the data of the other relation to all reducers. Of course since $r>s$, it makes sense to choose relation $R$ to partition. Thus
values of attribute $A$ are hashed to $k$ buckets, using a hash function $h$, and each tuple of relation $R$  with $B=b$ is sent to one reducer -- the one that corresponds to the bucket to which the value of the first argument of the tuple was hashed. The tuples of $S$ are sent to all the $k$ reducers.
Thus the number of tuples transferred from mappers to reducers is $ r + ks$.
\end{example}


The approach described above appears not only in Pig and Hive, but dates back to~\cite{WolfDY93}. The latter work, which looked at a conventional parallel implementation of join, rather than a MapReduce implementation, uses the same (non-optimal) strategy of choosing one side to partition and the other side to replicate.


In Example~\ref{2-way-ex} we show how we can do significantly better than the standard technique of Example~\ref{1-ex} and, thus, illustrating our technique.
%


\begin{example}
\label{2-way-ex}
We take again  the join $R(A,B) \bowtie S(B,C)$.
We partition the tuples of $R$ with $B=b$ into $x$ groups and we also partition the tuples of $S$ with $B=b$ into $y$ groups, where $xy = k$.
We use one of the $k$ reducers for each pair $(i, j)$ for a group $i$ from R and for a
group $j$ from $S$. Now we are going to
partition tuples from R and S, and we use hash functions $h_r$ and $ h_s$ to do the partitioning.
We send each tuple $(a,b)$ of $R$ to all reducers of the form $(i, q)$,
where $i=h_r(a)$
 is the group in which tuple $(a,b)$ belongs, and $q$ ranges over all $y$ groups. Similarly,  we send each tuple $(b,a)$ of $R$ to all reducers of the form $(q, i)$,
where $i=h_s(a)$
 is the group in which tuple $(b,a)$ belongs, and $q$ ranges over all $x$ groups.
Thus each tuple with $B=b$ from $R$ is sent to $y$ reducers, and each tuple with
$B=b$ from $S$ is sent to $x$ reducers. Hence the
communication cost is $ry + sx$. We can show
that by minimizing $ry + sx$ under
the constraint $xy = k$ we achieve communication cost equal to $2\sqrt{krs}$, which is always
less than what we found in Example~\ref{1-ex}, which was $ r + ks$.




\end{example}


%

\subsection{Our Setting}

We saw how to compute  the 2-way join  in Example~\ref{2-way-ex} for the tuples that have one HH (heavy hitter).
For this join, we  took two sets of keys:
\squishlisttwo
 \item The set of keys as presented   in Example~\ref{2-way-ex},
which send tuples with HH to a number of reducers in order to compute the join of tuples with HH.
\item The set of keys that send tuples without HH to a number of reducers in order to compute the join of tuples without HH. This second set of keys is formed as in a hash join.
\squishend
It is convenient to see these two sets of keys as corresponding to two joins which we call
{\em residual joins}, and which actually differ only on the subset of the data they are applied. One applies the original join on the data with HH and the
other  applies the original join on the data without HH.



In our setting, we assume a constraint that sets an upper bound $q$ on the size of the reducers. We use the Shares algorithm
to compute the shares for each attribute as a function of the total number of reducers, $k$. After that
we bound the number of inputs that are sent to each reducer by $q$ and we compute how many
reducers we must use for each residual join.
As the communication cost is increasing with the number of reducers, this is the best strategy compared to distributing all tuples for all residual joins to all reducers.

In the rest of the paper we will focus on
minimizing the communication cost as a function of the number of reducers, which we will denote $k$ through the paper. Since the Shares algorithm distributes tuples evenly to the reducers (the hash function we use sees to that), it is straightforward to enforce
the constraint that puts an upper bound on the size of each reducer, and thus compute the appropriate number of reducers needed.

%
%

\section{ Related Work}
The only other work that investigates skew when computing multiway joins in MapReduce is
\cite{BeameKS14,Magda15}.
In \cite{BeameKS14}, lower and upper bounds are given on the communication cost for algorithms that compute multiway joins in one round in share nothing architectures (it includes MapReduce but certain results therein capture more general models as well). For the upper bound, the Shares algorithm is shown to either meet the lower bound (when there is no skew) or offer a good upper bound in the presence of skew.
In both cases, the parameters of the map function (i.e., the shares -- see Section~\ref{shares-sec}
for details) are computed by a linear program which gives a solution to fractional edge packing of the hypergraph of the join. The main similarity of the algorithm we present in the present paper and the algorithm presented in \cite{BeameKS14} to handle skewed data is that, in both algorithms, the join to be computed is decomposed in a number of joins, called
residual joins. Each residual join is defined by a combination of heavy hitters and is applied on a different subset of the data. The combination of heavy hitters and the definition of a heavy hitter differ in the two papers, however.
\cite{Magda15} is a great paper which combines various known techniques (including those in  \cite{BeameKS14})
to investigate how various queries over datasets (including Twitter and Freebase
datasets) can be efficiently computed.


In the rest of this section, we review work either on 2-way joins with skew considerations or on multiway joins without skew considerations. We also review work on the Shares algorithm and recent developments on optimal serial algorithms for computing multiway joins.


A theoretical basis for a
family of techniques (including the Shares algorithm) to cope with skew by relating them to geometry is described in \cite{NgoRR13}.
In \cite{BeameKS13} it is proven that with high probability the  Shares algorithm\footnote{called HyperCube algorithm in this paper} distributes tuples evenly
on uniform databases (these are defined precisely in \cite{BeameKS13} to be databases which resemble the case of
random data). This class of databases include databases where all relations have the same size and there is no skew.




It was only a few years ago that an optimal serial algorithm to compute multiway joins was discovered \cite{NgoPRR12}. Previous serial algorithms could be significantly suboptimal when there was significant skew and/or there were many {\em dangling tuples} (tuples that do not contribute to the result).
In  \cite{NgoNRR14} a new algorithm is described that is able to satisfy stronger runtime guarantees than previous
join algorithms for data in indexed search trees.





Handling skew in MapReduce joins is considered also in the following papers.
 In \cite{Magda1,Magda2}, Skewtune is introduced as a system to mitigate skewness in real applications on Hadoop.
%
In \cite{HuangF14} and   \cite{TaoLX13}, multi-round   MapReduce algorithms  are considered with careful load balancing techniques.
%
%
 A Comparison of Join Algorithms for Log Processing in MapReduce is provided in  \cite{BlanasPERST10}, where skew is also discussed.
%



In \cite{HuaiCGHHOPYL014}, the authors significantly improve cluster resource utilization and runtime performance
of Hive by developing a highly optimized query planner
and a highly efficient query execution engine that handles skew as well.  Work in \cite{HuaiCGHHOPYL014} is done to facilitate users (as was Hive's original goal) to pose SQL queries on distributed computation frameworks by hiding from them the details of query execution. This is different from the goal of our
paper or the work in \cite{BeameKS14,Magda15}.


%

{\bf Multiway join in MapReduce without skew considerations}
Multiway join in MapReduce without skew considerations is examined in \cite{WuLMO11},
where a query optimization scheme is presented for MapReduce computational environments.
The query optimizer which is designed to generate an efficient query plan is based on  multiway join algorithms.
Another system implemented in  MapReduce and based on multiway joins is presented in \cite{JiangTC11}.

There are a few papers on theta-join in MapReduce (some addressing skew), including \cite{OkcanR11} and \cite{ZhangCW12} which focus on optimal distribution of data to the reducers for any theta function.
Efficient multi-round MapReduce algorithms for acyclic multiway joins are also developed recently  \cite{AfratiJRSU14,semih}.
In \cite{SakrLF13}, a comprehensive survey is provided for large scale data processing mechanisms based on
MapReduce.

\section{Shares Algorithm}
\label{shares-sec}
The algorithm is based on
a schema according to which we distribute the data to a given number of $k$
reducers. Each reducer is defined by a vector, where each component of the vector corresponds to an attribute.
The algorithm uses a number of independently chosen random hash functions $h_i$
one for
each attribute $X_i$. Each tuple is sent to a number of reducers depending on the value of $h_i$ for the
specific attribute $X_i$ in this tuple. If $X_i$ is not present in the tuple, then the tuple is sent to all
reducers for all $h_i$ values.
For an example, suppose we have the 3-way join $R_1(X_1,X_2) \bowtie
R_2(X_2,X_3) \bowtie R_3(X_3,X_1)$.
In this example each reducer is defined by a vector $(x,y,z)$.
A tuple $(a,b)$ of $R_1$ is sent to a number of reducers
and specifically to reducers
$(h_1(a), h_2(b), i)$ for all $i$. I.e., this tuple needs to be replicated a number of times, and specifically
in as many reducers as is the number of buckets into which $h_3$ hashes the values of attribute $X_3$.

When the hash function $h_i$ hashes the values of attribute $X_i$ to $x_i$ buckets, we say that the {\em share} of $X_i$ is $x_i$.
The communication cost is calculated to be, for each relation, the size of the relation times the replication
that is needed for each tuple of this relation. This replication can be calculated to be the product
of the shares of all the attributes that do not appear in the relation.
In order to keep the number of reducers equal to $k$, we need to calculate the shares so that their product is equal to $k$.

Thus, in our example, the communication cost is $r_1x_3+r_2x_1+r_3x_2$ and we must have $x_1x_2x_3=k$. (We denote the size of a relation $R_i$ by $r_i$.) Using the Lagrangean method (\cite{AfUl}), we find the
values that minimize the  cost expression: $x_1=(kr_1r_3/r_2^2)^{\frac{1}{3}}$, $x_2=(kr_1r_2/r_3^2)^{\frac{1}{3}}$ and $x_3=(kr_2r_3/r_1^2)^{\frac{1}{3}}$,
and thus the minimum communication is $3(kr_1r_2r_3)^{\frac{1}{3}}$.
We  give a more detailed example of the Lagrangean method in
 Section~\ref{dom-sec}.


\subsection{Dominance Relation}
\label{dom-sec}

An attribute $A$ is {\em dominated} by an attribute $B$ in the join if $B$ appears in all relations where $A$ appears.
It is shown  in \cite{AfUl} that if an attribute is dominated, then it does not get a share, or, in other words, its share is equal to 1. Now we give an example to illustrate the original Shares algoritm and the dominance relation.

\begin{example}
We repeat from \cite{AfUl} the example about a 3-way join. So, let $R(A,B)$, $S(B,C)$ and $T(C,D)$ be three binary relations whose join we want to compute, and let $r$, $s$ and $t$ be their sizes respectively.
First we observe that attribute $A$ is dominated by $B$ and $D$ is dominated by $C$, hence we do not include them in the communication cost expression since each gets share equal to 1.
Thus, the communication cost expression that we want to minimize is $ry+s+tx$, where $x$ is the number of shares for attribute $B$ and $y$ is the number of shares for attribute $C$.

We use the method of Lagrangean multipliers to solve and find the $x$ and $y$ that minimize the cost expression $ry+s+tx$ under the constraint $xy=k$. We begin with the equation
$ry + s + tx  - \lambda(xy - k) $,
take partial derivatives with respect to the two variables $x$ and $y$, and set the results equal to zero. We thus get that: (1) $r=\lambda x$, which implies $ry=\lambda xy=\lambda k$, and (2) $t=\lambda y$, which implies $tx =\lambda xy=\lambda k$.



\item


If we multiply (1) and (2) we get $rtxy=rt k=\lambda^{2} k^{2}$, which implies $\lambda=\sqrt{rt/k}$. From (1) we get $x=\sqrt{kr/t}$ and from (2) we get $y=\sqrt{kt/r}$. Thus the communication cost, which is $ry+tx$, is  equal to $\sqrt{2krt}$ when we substitute for $x$ and $y$. Hence, we  proved that $\sqrt{2krt}$ is the optimal communication cost.

Now how do we  hash values to the reducers?
We were given $k$ reducers.
A tuple $t=(u,v)$  of relation $S$
is sent only to one reducer, the reducer $(h_B(u), h_C(v))$ where $h_B$ and $h_C$ are the hash functions
that partition the values of attribute $B$ into $x$ buckets and the values of attribute $C$ into
$y$ buckets respectively. Now, a tuple $t=(u,v)$  of relation $R$ is sent to $y$ reducers, i.e., to all reducers with first component of their vector equal to $h_B(v)$. Similarly a tuple $t=(u,v)$  of relation $T$ is sent to $x$ reducers, i.e., to all reducers with second component of their vector equal to $h_C(u)$.

\end{example}

\section{overview of SharesSkew algorithm
}


The Shares algorithm fixes the number of reducers and optimizes under this constraint. In this paper, we do not
 fix the budget of reducers and try to apportion them among what could be an exponential number of different joins as in the Shares algorithm.  Instead we fix the reducer size $q$ (i.e., the number of input allowed in each reducer) and find how many reducers we should use for each residual join under this constraint.
 Thus,  $q$ will  determine the dimensions of the $k$-dimensional rectangle of reducers that we use for any special case involving HHs (make the dimensions big enough that the tuples from each relation are distributed so no reducer gets more than $q$ tuples in total).

\subsection{Partitioning Relations}
So after identifying HHs for each attribute that is not dominated, we partition each relation into a possibly exponential (in the number of attributes) number of pieces, depending on whether a tuple has no HH or which particular HH it has in each of its nondominated attributes (thus we have the {\em types} that are explained in detail in Section~\ref{com-expr-sec}).
Then, we consider in turn each combination of choices for all those nondominated attributes i.e.,  for each attribute  we have a type which is either a non-HH or a particular HH. Each combination defines one
residual join.

 Consider the example  $R(A,B) \bowtie  S(B,E,C) \bowtie T(C,D)$.  Suppose $B$ has HHs $b_1$ and $b_2$, while $C$ has HHs $c_1$, $c_2$, and $c_3$.  Then there are 12 combinations, depending on whether $B$ is $b_1$, $b_2$, or something else, and on whether $C$ is $c_1$, $c_2$, $c_3$, or something else.  $R$ is partitioned into three pieces, depending on whether $B$ is $b_1$, $b_2$, or something else.  $T$ is partitioned into 4 pieces, and $S$ is partitioned into 12 pieces, one for each of the 12 cases mentioned above
 (see some more details in Example~\ref{run2-ex}).

So for each combination of choices, we will join the parts of each relation that agree with that choice.
How we do this partitioning and  how we distribute tuples to reducers accordingly is described in  detail
Section~\ref{summary-sec}.


\subsection{Number of Reducers}
For each combination of choices, solving by Lagrangean multipliers, we get  optimal shares as a function of $k$, the total number of reducers for this combination of choices.  But we want to pick $k$ so no reducer gets more than $q$ tuples.  We compute the number of tuples that are expected to wind up at a reducer by dividing the communication cost (given as a function of the number of reducers, $k$) by the number of reducers $k$. This way, we get $k$ for each combination and get the shares for each attribute accordingly.

Once we do this for each combination of choices (i.e., residual join), and we get the proper $k$ as a function of $q$ for each, we can add those $k$'s for each residual join  to get the total number of reducers we need, as a function of $q$.
%
%

\section{SharesSkew Algorithm}
\label{ex-algo-sec}

This section contains the detailed description of our algorithm after we formally define residual joins.
In the end of the section we have two simple examples of running the algorithm, while, in next section,
we give a more elaborate and complete example.

\subsection{Definition of Residual Joins}
\label{com-expr-sec}

 For each attribute  $X$ we define a set $L_{X}$  of {\em types}:
\squishlisttwo
\item
If $X$ has no heavy hitter values, then $L_{X}$ comprises of only one type, $T_{-}$,  called the {\em ordinary type}.\footnote{Ordinary type represents all other values of attribute $X$, the ones that are not heavy hitters.}
\item If $X$ has $p$ values that are heavy hitters, then $L_{X}$ comprises  of $1+p$ types:
one type $T_b$ for each heavy hitter, $b$, of $X$, and one ordinary type $T_{-}$.
\squishend


\squishlisttwob
\item
A {\em combination of types}, $C_T$, is an element of the Cartesian product of the sets $L_{X_i}$, over all attributes $X_i, i=1,2,\ldots$, and defines a {\em residual join}.
 \squishend

We say that a tuple of relation $R$ is {\em relevant} to combination $C_T$ if it satisfies the constraints of
$C_T$.
Given a combination of types $C_T$, another combination $C'_T$ is a {\em subsumed} combination of $C_T$ if whenever $C_T$ and $C'_T$ disagree on a position (say that corresponds to attribute $B$), then the type of
$B$ in  $C_T$ in ordinary, the type of $B$ in $C'_T$ is non-ordinary and, for each relation $R$, the share $b$ of $B$ is less than $r/b_h$ where $r$ is the relevant size of relation $R$ and  $b_h$ is the number of tuples in $R$ where the specific HH of $B$ appears. We define the set of combinations that are considered by the algorithm to be the maximal set such that no combination in the set is subsumed by another combination in the set. For lack of space we do not include in the pseudocode how we find this set.

E.g., for the query in  Example~\ref{2-way-ex}, we consider two residual joins, one for type combination $C_T=\{ A:T_{-}, B:T_{-},  C:T_{-} \}$ (without HH) and one for type combination $C_T=\{ A:T_{-}, B:T_{b},  C:T_{-} \}$ (with HH). The other combinations are subsumed.

Each $C_T$ defines a {\em residual join} which is the join computed only on a subset of the data.
Specifically, if an attribute $X$ has ordinary type in the current $C_T$ we  exclude the tuples for which $X=HH$.
 If attribute $X$ is of type $T_b$ then we exclude (from all relations) the
tuples with value $X\neq b$.


\subsection{Description of the SharesSkew Algorithm}
\label{summary-sec}
We defined residual joins. Now
we need to define how to hash on $k$ reducers the relevant tuples for each residual join. As in the original Shares algorithm, we write the communication cost expression in terms of the shares variables for the attributes and then we minimize this expression under the constraint that the product of the shares is
equal to $k$. However, for each residual join we have a different cost expression, as we will explain in stage 3 below.

 The  SharesSkew Algorithm consists of the following four stages:
\squishlisttwo
\item[\textbf{Stage 1}] We form residual joins.
\item[\textbf{Stage 2}]  For each residual join, we form a set of keys by computing shares that minimize the cost expression for this residual join.
\item[\textbf{Stage 3}]  The cost expression for each residual join is the \emph{generic} cost expression of the original join where we have omitted the share variables
for attributes with HH -- in other words each such share is equal to 1. Moreover the size of each relation in the cost expression
is now equal to the number of tuples that satisfy the constraints of the specific residual join.
\item[\textbf{Stage 4}]  We distribute tuples according to the set  $K_J$ of keys we constructed for each residual join $J$, specifically: A tuple $i$ is hashed
according to the set of keys  $K_J$  if $i$ contains as values of HH attributes of $J$ the value that defines $J$.
\squishend

Below, we present the pseudocode for the SharesSkew algorithm. Stages 1-3 are implemented in \emph{PreMap} and stage 4 in the main \emph{Map} step.

\IncMargin{1em}
\begin{algorithm}
\SetAlgorithmName{Map Phase}{PreMap Step}{List of map tasks}
\SetKwFunction{GenericCost}{ConstructGenericJoinCostExpression}
\SetKwFunction{applyDominanceRule}{applyDominanceRule}
\SetKwFunction{solveLagrangean}{solveLagrangean}
\SetKwInOut{Input}{input}\SetKwInOut{Output}{output}
\Input{Relations schemas, \emph{HH[]}: HeavyHitters , \emph{J}: Join}
\Output{shares[]: Shares for the Combinations of HHs}
\BlankLine
$cost\leftarrow$ \GenericCost{}\;
\ForEach{combination $c$ of \emph{HHs}}{
	\ForEach{HH}{
    $cost_c \leftarrow cost_c[HH share=1]$ \;
    }
    \applyDominanceRule{$cost_{c}$} \;
    shares[$cost_{c}$] $\leftarrow$ \solveLagrangean{$cost_{c}$} \;
}

\caption{PreMap Step}\label{PreMap}
\end{algorithm}\DecMargin{1em}

\IncMargin{1em}
\begin{algorithm}
\SetAlgorithmName{Map Phase}{MainMap Step}{List of map tasks}
\SetKwFunction{GenericCost}{ConstructGenericJoinCostExpression}
\SetKwFunction{applyDominanceRule}{applyDominanceRule}
\SetKwFunction{solveLagrangean}{solveLagrangean}
\SetKwFunction{recursiveKeys}{recursiveKeys}
\SetKwInOut{Input}{input}
\SetKwInOut{Output}{output}
\Input{\emph{t}: tuple, \emph{HHs} list, attribute shares for each residual join, \emph{J}: Join, joinkey}
\Output{key-value pairs}

\BlankLine

\For{attribute Value $v$ of $t[i]$ in tuple $t$}{
	\For{Heavy Hitter ${hh_j}$ in $i$ position of $HH$s list}{
    \lIf{$v==hh_j$}{$C_t[i]=hh_j$ }
    }
}

\For{all HH combinations C that $C[i]==C_t[i]$}{
	\For{attribute in $joinkey$}{
    \If{$t[attribute]$ exists}{
    	\eIf{joinkey[attribute] exists}{
			match[attribute] $\leftarrow h$ \; }
         { match[attribute] $\leftarrow 1$ \; }
        }
    \ElseIf{$t[attribute]$ doesn't exist}{
    \If{joinkey[attribute] exists}{
			match[attribute] $\leftarrow r$ \; }
            }
    }

\For{$i$ in $match[]$}{
	\If{$match[i]$ == h}{
    	$base\_key[i]$ $\leftarrow$ $hash(t[i])$ \;
        }
    }
$key\_list$ = \recursiveKeys{$base\_key$, $n$:number of attributes} \;
\ForEach{key in $key\_list$}{
	emit($<key;value>$) \;
    }

}

\caption{Map Step}\label{mainmap}
\end{algorithm}\DecMargin{1em}

\begin{algorithm}
\SetAlgorithmName{Function}{Map Step}{List of map tasks}
\SetKwFunction{GenericCost}{ConstructGenericJoinCostExpression}
\SetKwFunction{applyDominanceRule}{applyDominanceRule}
\SetKwFunction{solveLagrangean}{solveLagrangean}
\SetKwFunction{recursiveKeys}{recursiveKeys}
\SetKwInOut{Input}{input}
\SetKwInOut{Output}{output}
\Input{\emph{$base\_key$}, \emph{n}: number of attributes, share sizes}
\Output{A list of keys}

\BlankLine
$recursive\_keys(base\_key, n)$: \;
\If{$n == -1$}{
	return key \;}
\If{$base\_key[n] == r$}{
	\For{$i\leftarrow 0$ \KwTo $share$}{
    	$key \leftarrow base\_key$ \;
        $key[n] = i$ \;
        $recursive\_keys(key, n-1)$ \;
	}
}
\Else{
        $recursive\_keys(key, n-1)$ \;
}

\caption{$recursive\_keys()$: Recursive Keys Construction}\label{mapkeyconstruction}
\end{algorithm}\DecMargin{1em}

The premap step is executed before the main \emph{Map step} on each mapper. In the input, $HH[]$ is a list of sets, where each set contains all of HHs of the i-th attribute. In line 1, the generic cost expression is constructed from the Join query and the relation schemas.
Then for each possible combination of HHs, i.e. for each \emph{residual join}, the share variable of each {HH} in the cost expression is set equal to $1$ (lines 3-5) and, in turn, the \emph{dominance rule} is applied to the residual cost expression (line 6). Finally, we solve for the shares that minimize that cost using the Lagrangean technique (line 7).

\begin{itemize}
\item In the main map step, we first identify which residual joins are compatible with the input tuple in hand (lines 1-5). The joinkey is implied by the shares, i.e. if an attribute has a share of 1 then it is not included in the joinkey, otherwise it is.
\item
Then, for all  residual joins (or \emph{HH combinations}), we mark with $h$ the attributes that are both present in the tuple and the joinkey (line 10) and we mark them with $1$  if they are \textbf{not} present in either  the tuple or the joinkey (line 12).
\item
Attributes that appear in the joinkey but not in the current tuple are to be replicated across reducers, thus marked with $r$ (line 17).
\item
We then construct a $base\_key$ by \emph{hashing} the values coming from attributes marked by $h$ (lines 21-25). In line 26, we \emph{recursively} construct (using the function $recursive\_keys()$) the set of keys which will determine the distribution of the current tuple to the reducers (lines 27-29).
\item
Thus, line 28 emits the set of key-value pairs for the particular residual join of the loop that started in line 6; this is what we denoted as $K_J$ in stage 4 above.
$recursive\_keys()$ builds each key in the key-set in a bottom-up fashion, by generating keys when it encounters attributes marked with $r$ (i.e. \emph{replicate} - lines 5-10).
\end{itemize}

 Observe that since the HH attributes do not get shares, the following is true:
\begin{itemize}
\item Each tuple is hashed to reducers according to the values of the non-HH attributes in this tuple.
\item If all attributes in the tuple are HHs then, this means that, in the current residual join, there is
only this tuple that participates (i.e., is relevant) from its relation, and it is hashed to all reducers.
\end{itemize}

\subsection{SharesSkew Algorithm for 2-way Join}
\label{Langrange}

Essentially what we suggested in Example~\ref{2-way-ex} can be summarized as follows. We decompose the 2-way join
into two  residual joins. They both compute the same query but on different data:
\begin{enumerate}
\item The first residual join computes the join on all tuples that do not contain the HH value.
The Shares algorithm for this join is trivial, we do not have replication of the tuples; hence
the communication cost is equal to the sum of sizes of the two relations (counting only the tuples
without HH).
\item The second residual join computes the join on only the tuples that contain the HH value of $B$.

\end{enumerate}
Let us see how the SharesSkew algorithm applies to 2-way join.
We first assume that all three attributes of the join are hashed according to their values.
Then we write the generic communication cost expression which is $ryz+sxz$.
For the second residual, we put the
share variable for attribute $B$ (here this is $z$) equal to 1. Hence we need
to minimize $ry+sx$ under the constraint that $xy=k$. This is what we did intuitively in Example~\ref{2-way-ex}.

\subsection{Example:  Combining Heavy Hitters in SharesSkew Algorithm}

The  two examples here explain what we do if an attribute
has more than one heavy hitter or when there are several attributes with heavy hitters.

\begin{example}
\label{more-hh-ex}
Suppose in the 2-way join $J=R(A,B) \bowtie S(B,C)$ we have two  heavy hitters for attribute $B$ (say $b_1$ and $b_2$).  Then we have three residual joins: one without HH, one with only the tuples with $B=b_1$
and a third one with only the tuples with $B=b_2$.

%
\end{example}


Now an example when we have several attributes with possibly more than one HH each.
\begin{example}
\label{run2-ex}
We take as our running example the  3-way join:~ $J=R(A,B)\bowtie S(B,E,C) \bowtie T(C,D)$.
Suppose attribute $B$ has two HHs,  $B=b_1$ and $B=b_2$ and attribute $C$ has one HH,
and $C=c_1$. Now we have six residual joins: 1) no HH in all attributes, 2) $C$ with HH but no
HH in attribute $B$, 3) $B$ with HH  $B=b_1$ and no HH in $C$, 4) $B$ with HH  $B=b_2$ and no HH in $C$,
5) $B$ with HH  $B=b_1$ and  HH in $C$, 6) $B$ with HH  $B=b_2$ and HH in $C$.

\end{example}


\section{Applying the SharesSkew Algorithm on an Example}
In this section we include an elaborate and detailed example for SharesSkew.
\begin{example}
\label{run1-ex}
We take again Example~\ref{run2-ex}.
We repeat here. We want to compute the  3-way join:~ $J=R(A,B)\bowtie S(B,E,C) \bowtie T(C,D)$.
Suppose attribute $B$ has two HHs,  $B=b_1$ and $B=b_2$ and attribute $C$ has one HH,
and $C=c_1$. Thus attribute $B$ has three types, $T_{-}$, $T_{b_1}$ and $T_{b_2}$, attribute
$C$ has two types, $T_{-}$ and $T_{c_1}$ and the rest of the attributes have a single type, $T_{-}$.
Thus we have $3\times 2=6 $ residual joins, one for each combination.
By $r,s,t$ we denote the sizes of the relations that are {\em relevant} in each residual join, i.e., the number of tuples from each relation that
contribute in the particular residual join.  We list the residual joins:
\squishlisttwo
\item All attributes  of type $T_{-}$. Here  $r$ is the number of only those tuples of relation $R$ for which
 $B\neq b_1$ and $B\neq b_2$, $s$ is the number of only those tuples of relation $S$ for which
 $B\neq b_1$ and $B\neq b_2$ and $C\neq c_1$, and $t$ is the number of those  tuples in relation $T$ for which $C\neq c_1$.
 \item All attributes of type $T_{-}$, except  $B$ of type $T_{b_1}$.
  In this case  $r$ is the number  of only those tuples in relation $R$ for which
 $B= b_1$,  $s$ is the number of only those tuples in relation $S$ for which $B= b_1$ and $C\neq c_1$, and $t$ is the number of those  tuples in relation $T$ for which $C\neq c_1$.
\item All attributes  of type $T_{-}$, except  $B$  of type $T_{b_2}$.
The analysis almost same as the case above with the only difference that we have a different HH for $B$, hence different sizes for
  relations $R$ and  $S$.
 \item All attributes  of type $T_{-}$, except  $C$ of type $T_{c_1}$.
 The treatment is similar as in (2) above for $B$.
 \item All attributes a of type $T_{-}$, except  $B$ of type $T_{b_1}$ and  $C$  of type $T_{c_1}$.
  In this case  $r$ is the number of only those tuples of relation $R$ for which
 $B= b_1$, $s$ is the number of only those tuples of relation $S$ for which
 $B= b_1$ and $C=c_1$, and $t$ is the number of those  tuples in relation $T$ for which $C= c_1$.
 \item All attributes  of type $T_{-}$, except  $B$ of type $T_{b_2}$ and  $C$  of type $T_{c_1}$.
 The analysis is analogous to the case (5) above.
\squishend
\end{example}
Each residual join is treated by the Shares algorithm as a separate join and a set of keys is defined that will be used to hash
each tuple as follows: A tuple $t$ of relation $R_j$ is sent to reducers of combination $C_T$ only if the values of the tuple satisfy the constraints of $C_T$ as concerns values of HH. We give an example:
\begin{example}
\label{run111-ex}
We continue from Example~\ref{run1-ex}.
Each tuple is sent to a number of reducers according to the keys created for each
residual join. E.g., a tuple $t$ from relation $R$ is sent to reducers as follows:
\squishlisttwo
\item If $t$ has $B=b_1$ then it is sent to reducers created in items (2) and (5)  in Example~\ref{run1-ex}.
\item If $t$ has $B\neq b_1$  and $B\neq b_2$ then it is sent to reducers created in items (1) and (4) in Example~\ref{run1-ex}.
\item If $t$ has $B=b_2$ then it is sent to reducers created in items  (3) and (6) in Example~\ref{run1-ex}.
\squishend

\end{example}


In Example~\ref{run1-ex} we showed how to construct the residual joins and in Example~\ref{run111-ex} we showed how to distribute tuples. Now we are going to show how to write the cost expression for each residual join and compute the shares.


\begin{example}
\label{cost-ex}
We continue from Example~\ref{run1-ex} for the same HH as there.
 Remember by $a,b,c,d,e$ we denote the shares for each attribute $A,B,C,D,E$ respectively and by $r,s,t$ we denote the sizes of the relations that are {\em relevant} in each residual join, i.e., the number of tuples from each relation that
contribute in the particular residual join.
We always start with the generic cost expression for the original join, $r   cde +   s  ad +  t abe  $, and then simplify accordingly.
We list the cost expression for every residual join (and in the same order as) in Example~\ref{run1-ex}:
\squishlisttwo
 \item Here all attributes are ordinary, so we simplifly the relation by observing that
 $A$ is dominated by $B$ and $D$ is dominated by $C$, hence $a=1$ and $d=1$ and the expression is:
$r   c +   s   +  t b  $.

 \item Here only $B$ is a non-ordinary attribute, hence $b=1$ and then, from the remaining attributes
  $D$ and $E$ are  dominated by $C$,  hence $d=1$and $e=1$, and the expression is:
 $r   c +   s  a +  t a  $
 \item All attributes are of type $T_{-}$, except attribute $B$ which is of type $T_{b_2}$. The analysis almost same as the case above with the only difference that we have a different HH for $B$, hence different sizes for
 relevant relations. Thus, the expression is
    $r   c +   s  a +  t a  $, i.e., same as above, only the sizes of the relations will be different.
 \item
 All attributes are of type $T_{-}$, except attribute $C$ which is of type $T_{c_1}$.
 Hence $c=1$. From the rest of the attributes, $A$ and $E$ are dominated by $B$.
 Thus, the expression is
 $r   d +   s  d +  t b  $.
%
 \item Here we set both $b=1$ and $c=1$ and this gives us $r   de +   s  ad +  t ae  $.

\item The expression is $r   de +   s  ad +  t ae  $, i.e., same as above, only the sizes of the relations will be different.
\squishend

\end{example}

\section{Efficiency of  SharesSkew }
\label{proof-opt-sec}

In this section, we will explain why the SharesSkew algorithm is expected to work efficiently.
In SharesSkew, we designated arbitrarily no shares to the HH attributes  (i.e., their share =1). In this section we will show that this can be seen as an algorithm that runs on random data (instead of data with HH),
and thus it has the good properties of the Shares algorithm.
We first begin with 2-way join and then we explain for any multiway join.
In the end we give a lower bound for the 2-way join which shows that, in this case, SharesSkew is optimal.

\subsection{2-way join}

 To see that the  method we presented in Example~\ref{2-way-ex} is actually based on the Shares algorithm
 (and to be able to extend it for more than one HH), we think as follows: We replace
 each tuple of relation $R$ with a tuple where $B$ has distinct fresh values
$b_1,b_2,\ldots $ and the same for the tuples of relation $S$ with $B$ having values $b'_1,b'_2,\ldots$.
Now we can apply the Shares algorithm to find the shares and distribute the tuples to reducers normally. The only
problem with this plan is that the output will be empty because we have chosen the $b_i$s and $b'_i$s
to be all distinct.
This problem however has an easy solution, because, we can keep this replacement
at the conceptual level, in order to create a HH-free join and be able to apply the Shares algorithm
and compute the shares optimally. When we transfer the tuples to the reducers, however, we transfer the
original tuples and thus, we produce the desired output. We formalize this thought for 2-way join in the next paragraph
  ( Section~\ref{proof-opt-sec} extends for any multiway join).

First we introduce two {\em auxiliary attributes} $B_R$ and $B_S$ and an {\em auxiliary relation} $R_{aux} (B_R,B_S)$.\footnote{The auxiliary attributes and relation are
only used in the conceptual level, as we will discuss in more detail in Section~\ref{proof-opt-sec}.}
Suppose we have the join $R(A,B_R) \bowtie R_{aux} (B_R,B_S)\bowtie S(B_S,C)$ to compute on three relations, where relation $R(A,B_R) $ contains all tuples
of relation $R(A,B) $ with $B$ being the HH but we have replaced the HH value with a fresh value,
different for each tuple (say set $B_1$ consists of all these values). Similarly,  for each tuple of relation $S(B,C) $ with $B$ being the HH  we have replaced the HH value with a fresh value,
different for each tuple (say set $B_2$ consists of all these values -- $B_1$ and $B_2$ are disjoint).
Also let relation $R_{aux} $ consist of the Cartesian product of $B_1$ and $B_2$. The 3-way join on this data is computed in an almost isomorphic way as the 2-way on the tuples with HH, i.e., same number of
pairs of tuples from $R$ and $S$ joining and same size of result. However, the 3-way join is a join without skew, so we can use the original Shares algorithm. Then the communication cost expression is
$rcb_s + r_{aux} ac +sab_r$ (where $c,b_s,a$ and $b_r$ are the shares for attributes $C,B_S,A$ and $B_R$ respectively). Here the auxiliary relation is not actually transferred; hence we can drop
from the cost expression the middle term. Since we dropped this term, it is as if we have a cost
expression for the join $R(A,B_R) \bowtie S(B_S,C)$.

Now, considering the dominance relation in the join $R(A,B_R) \bowtie S(B_S,C)$, we have a choice: either to choose that $A$ dominates $B_R$ or $B_R$ dominates $A$. We choose the former,
hence $b_r=1$. Similarly we take $b_s=1$. Hence the cost expression now is $rc+sa$.
Thus we have arrived at the same algorithm to compute 2-way join with skew as the one we developed intuitively in
Example~\ref{2-way-ex} and with the same communication cost.  Of course the values without skew are brought together with a different set of keys, since the  shares have values $\neq 1$ in different attributes.
However, if there are, say four HH on attribute $B$, then we can compute all four as a single multiway join
as we will explain in Section~\ref{proof-opt-sec}.

\subsection{Any Multiway Join}





The structure we use in this section is conceptual of course.
 In practice, we do not materialize  any of the auxiliary relations or attributes.


%



The conceptual structure in the general case is as follows:
 For each combination of types, $C_T$,  we define a HH-free residual join (it is HH-free by construction):
\squishlisttwo
\item If attribute $X$  has non-ordinary type  in $C_T$ then:\\
--We introduce a number of
auxiliary attributes, one auxiliary attribute  for each relation $R_j$ where  attribute $X$ appears.  We denote the auxiliary
attribute for relation $R_j$ by  $X_{R_j}$.\\
--In the schema of each relation $R_j$ where $X$ appears, we replace  $X$ with  attribute $X_{R_j}$.

\item We form the residual join $J'$ for $C_T$ by   adding to original join new relations as follows: one relation, $R^{X}_{aux}$,
for each attribute $X$ which is not
of ordinary type. The schema of~ that~ relation~ consists of the attributes
$X_{R_j}$  for each $j$ such that $X$ is an attribute of $R_j$.
\squishend


Now we apply this modified join $J'$ on the following database $D'$ that is constructed from the given database $D$ as follows:
\squishlisttwo
\item For each HH (in the current residual join) in a tuple $i$ of relation $S$ we do: Suppose
 value $a$ is a heavy hitter, then we replace $a$ with $a.i.S$ in tuple $i$.  We denote the set of all $a.i.S$'s
 by $A_S$.\footnote{Of course, we do not include any more tuples in $S$.}
\item We form each auxiliary relation by populating  it with the cartesian product of sets $A_S$, one set for each relation where $a$ is a HH in the current residual join.
\squishend

\squishlisttwob
\item {\em Observation 1:} Database $D'$ now has no heavy hitters.
\squishend
\begin{example}
Thus if, in the database $D$, relation $R$ is\\ $\{ (1,2),(3,2),(4,2)\}$ and $S$ is  $\{ (2,5),(2,6)\}$
then, in the database $D'$ we have (assuming $B=2$ qualifies for HH):\\\\
  $R(A,B_R) $ is $\{ (1,2.1.R),(3,2.3.R),(4,2.4.R)\}$.\\
$S(B_S,C)$ is  $\{ (2.5.S,5),(2.6.S,6)\}$.\\
(I.e., we conveniently identify the tuple of $R$ with the value of its first argument and the tuple of $S$ with the value of its second argument.)\\
The auxiliary relation $R_{aux}(B_R,B_S)$ is :\\ $\{ (2.1.R,2.5.S),(2.3.R,2.5.S),(2.4.R,2.5.S),$\\ $(2.1.R,2.6.S),(2.3.R,2.6.S),(2.4.R,2.6.S)\}$
\end{example}

\squishlisttwob
\item {\em Observation 2:} There is a tuple $i=(a_1,a_2,\ldots, d_1, d_2, \ldots)$ in relation $S$ in $D$ with $d_1,d_2,\ldots$ being the HH iff there is a {\em corresponding} tuple $i=(a_1,a_2,\ldots, d_1.i.S, d_2.i.S, \ldots)$ in $D'$. Hence, in the presence of the auixialiry relations too, a number of tuples form join $J$ in  $D$ iff the corresponding tuples form join $J$  in $D'$.
\squishend

{\bf Simplifying the Cost Expression}
Finally, in this subsection, we show that the cost expression for each residual join for the SharesSkew algorithm is indeed the one that computes the number of tuples transferred from the mappers to the reducers according to the schema that distributes the tuples. Hence by minimizing this expression we find the optimal solution as regards communication cost.

First we observe that the property of the dominance relation allows us to write the cost expression for each residual join
in a simple manner. We use the theorem:


\begin{theorem}
The share of each auxiliary attribute is equal to 1 in the optimum solution.
\end{theorem}

Thus we established that:
\squishlisttwob
\item Each tuple is hashed to reducers according to the values of the non-HH attributes in this tuple.
\squishend

\subsection{Lower Bound for 2-way Join}
\label{lb-sec}
Here we prove that the solution in Example \ref{2-way-ex} is optimal because it meets the lower bound that we compute in this section.

Suppose $r_q$ and $r_s$ is the number of tuples in each reducer from relations $R$ and $S$ respectively.
Suppose $r_q =\xi s_q$. Let communication cost be denoted by $c$. Then $c=k(r_q+s_q)=ks_q(1+\xi )$.
The total output is $rs$ and the output from all reducers is $kr_q s_q= k\xi s^2_q$. Thus we have the inequality:
$$rs\leq k\xi s^2_q$$
Hence
$\sqrt{ rs/(k\xi)}\leq  s_q $
Thus from the first equation above about the cost we have:\\
$c \geq k (1+\xi ) \sqrt{ rs/(k\xi)}$
or
$c \geq  (1+\xi ) \sqrt{ krs/\xi}$\\
But for all $\xi$ we have $(1+\xi )/ \sqrt{\xi }>2$
Hence $$ c\geq 2\sqrt{krs}$$

\section{Chain joins and Symmetric Joins}

In this section we provide closed forms for attribute shares and communication cost for chain joins for the
SharesSkew algorithm. Then we argue that there are classes of multiway joins (the symmetric joins) where Shares behaves
well enough on heavy skew. We give closed forms for computing shares and communication cost
for symmetric joins.

There is already formal (probabilistic) evidence \cite{BeameKS13} that the Shares algorithm behaves almost
optimally for certain joins even in the presence of skew. It is natural, therefore, to try to identify multiway joins for which Shares itself can handle skew. We provide such a family of joins here, the {\em symmetric joins}.
Moreover, we compare the performance of Shares algorithm on chain joins as well to illustrate the
reason of the difference. For chain joins, the communication cost of the Shares algorithm is high and
it increases with the number of relations added to the join. For symmetric joins, however, the communication cost of the Shares algorithm decreases with the number of relations in the join. Moreover, for symmetric joins, SharesSkew does not significantly improve performance compared to Shares -- since Shares already achieves close to optimal communication cost as we will show here.

\subsection{Chain Joins: all relations equal size}
We begin with chain joins which are joins of the form:
$$R_1(A_0,A_1)\bowtie R_2(A_1,A_2) \bowtie \ldots \bowtie R_n(A_{n-1},A_n)$$
When some attributes are HH in a specific residual join, then the cost expression as analyzed in
\cite{AfUl} changes because (as we already explained) the HH attributes take share equal to 1.
However, we make an observation here that leads to a trick that allows us to get closed forms
for the shares even for this different cost expression.

So, here we make the observation that each attribute in the chain that is a HH may be seen as dividing the chain in two parts (if there is only one HH) because in the cost expression the shares of the HH is equal to 1, so the expression can be viewed as the sum of two expressions each  for one subchain. We can, thus, use the formula from
\cite{AfUl} that gives the shares that maximize each subexpression, given that it uses $k_i, i=1,2$ reducers.
Now we need to minimize taking into account that $k_1k_2=k$.

In the general case, where we assume we have $m$ subchains (because we have $m-1$ HH) the constraint is $\Pi_{1}^m k_i=k$.
Each subexpression has communication cost (derived from \cite{AfUl}):
$$ r n_i k_i^{(n_i-2)/n_i}$$
where $r$ is the size of each relation and $n_i$ is the number of relations in the $i$-th (sub)join.
We want to minimize the:
$$\sum  r n_i k_i^{(n_i-2)/n_i}$$
By taking the Lagrangean, we find that
$$(n_1-2)  k_1^{(n_1-2)/n_1}=(n_i-2)k_i^{(n_i-2)/n_i}$$
Hence,
$$k_i=(\frac{n_1-2}{n_i-2})^{n_i/(n_i-2)}k_1^{n_i(n_1-2)/n_1(n_i-2)}$$
Hence, by multiplying all, and since $k_1k_2\ldots k_m=k$, we can get the $k_1$ and from it all the $k_i$'s.



Finally, the shares for each attribute that belongs in the subchain $i$ and is not in odd position within the subchain is $k_i^{1/n_i}$  otherwise it is =1 (as we mentioned, these calculated in \cite{AfUl}).

These calculations hold when all subchains have even length. For odd length, the case is similar but a little more tedious.

\subsection{Chain Joins: Arbitrary relation sizes}
\label{same-size-chain-sec}
First we compute the optimal communication cost for chain joins without HH (it is not included in \cite{AfUl}).
In this case, the cost expression from \cite{AfUl} is shown to be a sum of terms where each term $\tau _i $
is:
$\tau _1= r_1k/a_1$, $\tau _n= r_n k/a_{n-1}$ and
$\tau _i= r_i k/a_{i-1}a_i$, where $a_i$ is the share for attribute $A_i$.
Solving the Lagrangean obtains that all odd terms are equal to each other and the same for all even terms.
In particular, the equalities for the
even-$n$ case have the form

\begin{center}\begin{tabular}{l}
$\frac{r_1}{a_1} = \frac{r_3}{a_2a_3} = \frac{r_5}{a_4a_5} =
\cdots =
\frac{r_{n-1}}{a_{n-2}a_{n-1}}$\\
\\
$\frac{r_2}{a_1a_2} = \frac{r_4}{a_3a_4} = \cdots =
\frac{r_{n-2}}{a_{n-3}a_{n-2}} =
\frac{r_n}{a_{n-1}}$\\
\end{tabular}\end{center}

Thus, setting $\tau _i= r_i k/a_{i-1}a_i=\lambda _1$ for odd terms and $\tau _i= r_i k/a_{i-1}a_i=\lambda _2$ for even terms, and multiplying all odd terms to get get $\lambda _1^{n/2}$ and all even terms to get
$\lambda _2^{n/2}$ we observe that: the denominator of the left hand side (after this multiplication)
is the product of all $a_i$'s, hence it is equal to $k$.
Thus we get that
$$\lambda _1 =  k^{1-2/n} (r_1r_3r_5\ldots)^{2/n}$$
$$\lambda _2 =  k^{1-2/n} (r_2r_4r_6\ldots)^{2/n}$$
Thus the communication cost is:
$$cost=n/2(\lambda _1 +\lambda _2)=$$  $$n/2 \times k^{(n-2)/n}((r_1r_3r_5\ldots)^{2/n}+(r_2r_4r_6\ldots)^{2/n})$$

Now, for finding the optimal shares for chains with HH, exactly the same calculations as in Subsection \ref{same-size-chain-sec} apply; i.e., instead of multiplying the $k^{(n_i-2)/n_i} $ by  $n_i$ in the cost expression of each subjoin $i$ we multiply it by
 $$n/2 \times ((r_1r_3r_5\ldots)^{2/n}+(r_2r_4r_6\ldots)^{2/n})$$

\subsection{Symmetric Joins}
We define {\em symmetric} joins to be the joins whose associated hypergraph has adjacency matrix which a)
is $r$-diagonal in the sense that in the $i$-th row contains 1 in $i$ through $i+d$ (for a given $d$ for this join) entry and 0 otherwise, where $i+d$ is mod-$m$ where  $m$ is the length of a row and b) contains
$m+d$ rows where $m$ is the number of columns.

Thus properties of symmetric joins include:
\begin{itemize}
\item All relations have the same arity.
\item Each attribute appears in exactly $d$ relations.
\item Each subset of attributes of size $d$ appears in exactly one relation.
\end{itemize}
%

Our goal is to give a closed-form expression for communication cost for symmetric joins.

We shall first analyze the case where all relations are of equal
size.  That case involves considerably simpler algebraic
expressions, yet  serves to introduce
the calculations used in the general case, without obscuring the idea
behind the algebra.

With techniques very similar to chain joins as in \cite{AfUl} it is easy to get the Lagrangean equations for this case:

\begin{center}\begin{tabular}{l}
$\tau_{d+1} + \tau_{d+2} +\cdots+ \tau_n = \lambda k$\\
$\tau_1 + \tau_{d+2} +\cdots+ \tau_n = \lambda k$\\
$\tau_1 + \tau_2 + \tau_{d+3} +\cdots+ \tau_n = \lambda k$\\
$\tau_1 + \tau_2 + \tau_3 + \tau_{d+4} +\cdots+ \tau_n = \lambda k$\\
$~~~~~\ldots$\\

\end{tabular}\end{center}
These equations imply the following $d$ groups of equalities:

\begin{center}\begin{tabular}{l}
$\tau_1=\tau_{d+1}=\tau_{2d+1}=\cdots$\\
$\tau_2=\tau_{d+2}=\tau_{2d+2}=\cdots$\\
$\tau_3=\tau_{d+3}=\tau_{2d+3}=\cdots$\\
$~~~~~\ldots$\\
$\tau_d=\tau_{2d}=\tau_{3d}=\cdots$\\
\end{tabular}\end{center}
We shall henceforth look
only for values of the share variables (and give closed form solution to the communication cost)  that satisfy the $d$ groups of equalities. After we do the math we get the theorem:


\begin{theorem}
 For each join in the  class of symmetric joins with $n$ relations and $d$ attributes in each relation, the Shares algorithm  has communication cost equal to
 $$n_dk^{{1-\frac{d}{n}}}\sum_{all~S}\Big( \Pi _{i \in S} r_i \Big)^{1/n_d}$$
 where $n_d$ is the smallest integer such that $n$ divides $dn_d$, and each
 $S$ is the subset of relations $\{R_j, R_{j+d},R_{j+2d},\ldots, R_{j+dn_d}\}$, $j=1,2, \ldots$.
 \end{theorem}

\vspace{-0.1in}
 After the analysis in this section, we make a very important observation: There is a large class of symmetric joins with very low communication cost without having to take into account HH. Just observe the  communication cost which is proportional to $k^{1-\frac{d}{n}}$. In this case if $d$ is close to $n$ then
 this cost is almost optimal, e.g., for $d=n-3$ this cost is proportional to  $k^{\frac{3}{n}}$.
 In contrast, chain joins have high communication cost since it is proportional to $k^\frac{n-2}{n}$.

 Finally, it remains to explain how to find the share for each attribute. We observe that $\tau_i/\tau_{i+1}$ is equal to $a_{i-1}/a_{a+d-1}$. Since we have calculated the $\tau$'s, we can derive, thus, a simple linear system of equations which we can solve to find the $a$'s.

%

\subsection{Multi Rounds of MapReduce}
After the above analysis, it does not come as a surprise that two rounds (or more) of MapReduce
can have significantly lower communication cost than one round. Imagine the following scenario: There are two parts in the join (loosely joined with one another): a) a part with HH and symmetric which falls in the subclass of previous section with resiliency to skew and  b) a part such as a chain join without HH. Now, the communication cost of the chain join is high compared to the some of the costs we showed in previous section. If we use one round, then we will have
to distribute for each residual join the same tuples that are input to the schema of the first part of our join.
However, if we join the two parts separately we will have the following benefits: a) the first part will
run with low communication cost, b) the final join of the outputs of the two parts can be done with
minimum communication, thus c) the important overhead to the communication cost will come only from the second part of the join (the chain subjoin), which we could not avoid it anyway.

\section{Experiments}

In this Section we provide two sets of experiments to showcase the advantages of using SharesSkew algorithm as compared to Shares algorithm in the presence of skew in the data.

\subsection{2-way join}
\label{ssec:exp1}
For this comparison we consider the 2-way join $R(A,B) \bowtie S(B,C)$. The numbers of tuples populating $R$ and $S$ are $10^6$ and $10^5$, accordingly, and they contain a single HH which appears in 10\% of the tuples. Each tuple's size varies from a few bytes to 5KBytes.

This experiments bundle was set up as follows. We used an \texttt{Amazon Web Services EC2} cluster consisted of 16 \texttt{c3.2xlarge} instances running Ubuntu Linux 14.04 LTS. Each instance had 8~\texttt{vCPU}s with 15GB of memory and 80GB of SSD storage each, amounting to a total of 64~\texttt{vCPU}s.

\begin{figure}[]
 \centering
\subfigure[Shuffle time comparison]{%
\includegraphics[scale=0.45]{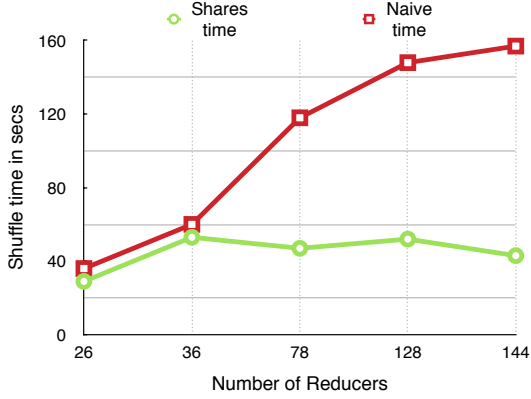}
\label{fig:subfigure1}}




\subfigure[Shuffled tuples for the SharesSkew and Shares algorithms  compared to the $2\sqrt{krt}$]{%
\includegraphics[scale=0.45]{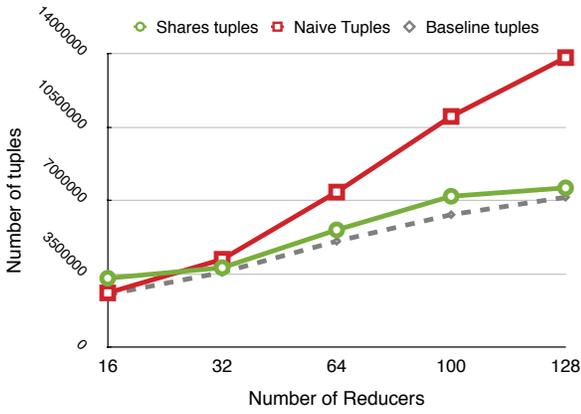}
\label{fig:subfigure2}}


\caption{Comparing shuffle and reduce times between Shares and Naive algorithm for $R(A,B) \bowtie S(B,C)$ }
\label{fig:figure}
\end{figure}

\begin{figure}[]

\includegraphics[scale=0.45]{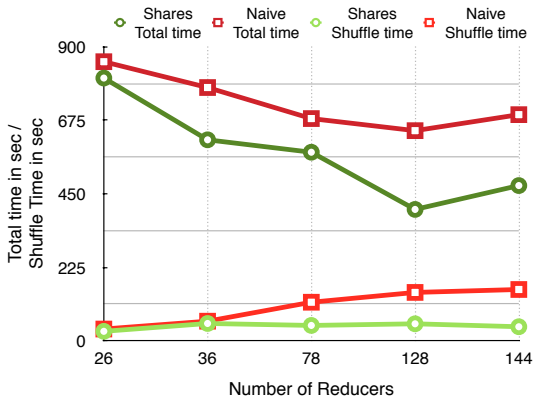}

\caption{Comparing compute times between Shares and Naive algorithm for $R(A,B) \bowtie S(B,C)$ }
\label{fig:figure2}

\end{figure}

In Figure~\ref{fig:subfigure1}, we show the shuffle time for the naive algorithm (as in Example~\ref{1-ex}) and the
SharesSkew algorithm (as in Example~\ref{2-way-ex}). In Figure~\ref{fig:subfigure2} we show the total time of these
algorithms and we have included the shuffle time again to compare what the percentage
of shuffle vs. total time is. As expected from our analysis, the shuffle time of the naive
algorithm is considerably larger than the SharesSkew algorithm. Moreover, the shuffle time
is a large enough percentage of the total time which will affect performance especially if
the size of the data is of the order of TB or larger and we have use a number of reducers
in the order of thousands. Finally, we have included Figure~\ref{fig:figure2} which shows the number of
tuples transferred from the mappers to the reducers. This is proportional to the shuffle time,
but we use it here to compare with our theoretical finding, which says that the
number of tuples transferred for SharesSkew is proportional to the square root of the number
of reducers -- this is the dotted line that we have drawn there to show that it is followed
in the experiments too.

\subsection{3-way join}

In this second part, we consider the 3-way join $R(A,B) \bowtie S(B,E,C) \bowtie T(C,D)$. Each relation in the join has $10^5$ tuples, and, in this case, attribute $B$ has two HH values, $b_1$ and $b_2$, and attribute $C$ has one, $C=c_1$, as in example~\ref{run2-ex}. These HHs accounted for 10\% of the total input.\\

For this experiment we used a Hadoop 2.6.0 cluster, consisting of 312 vCores with a a total of 2.2TB RAM spread across 40 nodes.

In Figure~\ref{fig:figure2}, we show the reduce times for the SharesSkew algorithm in the presence of skew compared to the Shares reduce time for input without skew. We also show the shuffle time for the SharesSkew algorithm.
The shuffle time for SharesSkew is considerably larger than the shuffle time of Shares (which is so small that is out of scale) because Shares sends the HH tuples to the same reducer. Observe that this is not the case with the naive algorithm in Section~\ref{ssec:exp1} which distributes HH tuples to reducers but does not do so optimally, whereas SharesSkew distributes them (in order to actually create a Cartesian product).
On the other hand, the reduce time of Shares on skewed data is very large and out of scale here.
We also run Shares on data without skew to demonstrate that SharesSkew achieves  performance on skewed data close to the performance that
Shares achieves on non-skewed data.

\begin{figure}[]

\includegraphics[scale=0.45]{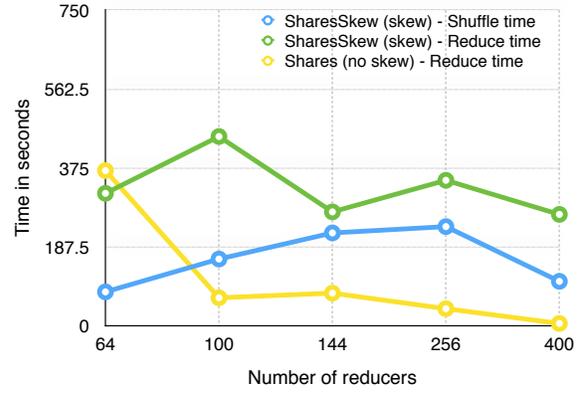}
\label{fig:subfigure5}

\caption{Comparing shuffle and reduce time for Shares and SharesSkew for $R(A,B) \bowtie S(B,E,C) \bowtie T(C,D)$}

\end{figure}

\subsection{Discussion}

As shown by the theoretical analysis and verified by the experiments, the SharesSkew algorithm retains the good properties of Shares algorithm on skewed data, i.e.:
\begin{itemize}
\item It distributes the tuples evenly to the reducers. Hence its performance scales with the number
of reducers.
\item Its performance does not depend on how much skew we have in the data. E.g., in Section~\ref{ssec:exp1},
we have 100 percent skew (we only include tuples with one HH).
\item It is particularly useful when we have large tuples (e.g., that may contain images) where the shuffle
time increases considerably. In our experiments the tuples were not very large.
\end{itemize}

\section{Conclusions}
We presented a MapReduce algorithm (SharesSkew) for multiway joins. The algorithm minimizes communication cost  and is tested with experiments which testify that the wall-clock time is significantly affected by the  communication
cost. In our experiments, we also compared the performance of SharesSkew to the performance of Shares
algorithm to show that SharesSkew has the good properties of Shares.

We also investigated the performance of SharesSkew algorithm in more detail as concerns how it performs in special classes of multiway joins.  We have shown
 that a) there exist multiway joins that Shares does almost as well as SharesSkew does even in the presence of skew and b) there exists a class of  multiway joins where, Shares has very high communication cost even for random data. This leads to open problems that should consider
 multi-rounds of MapReduce for certain classes of joins.


\bibliographystyle{abbrv}
\bibliography{skew}  
%
%

\end{document}